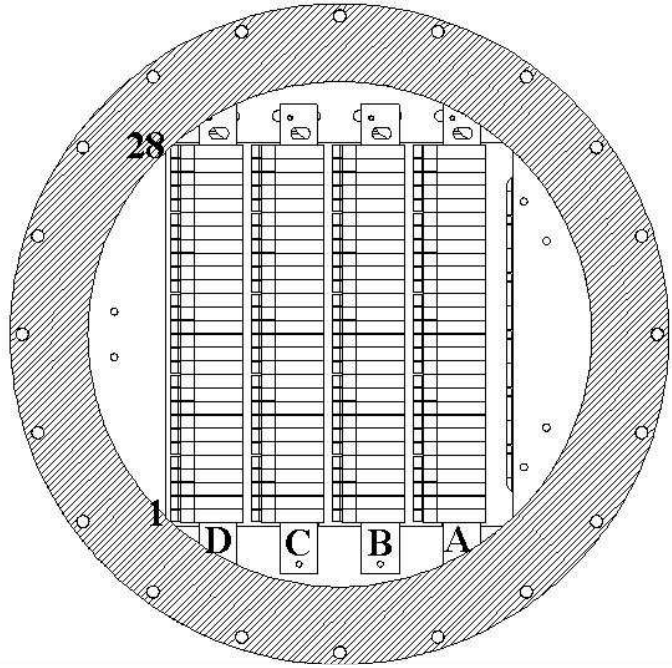

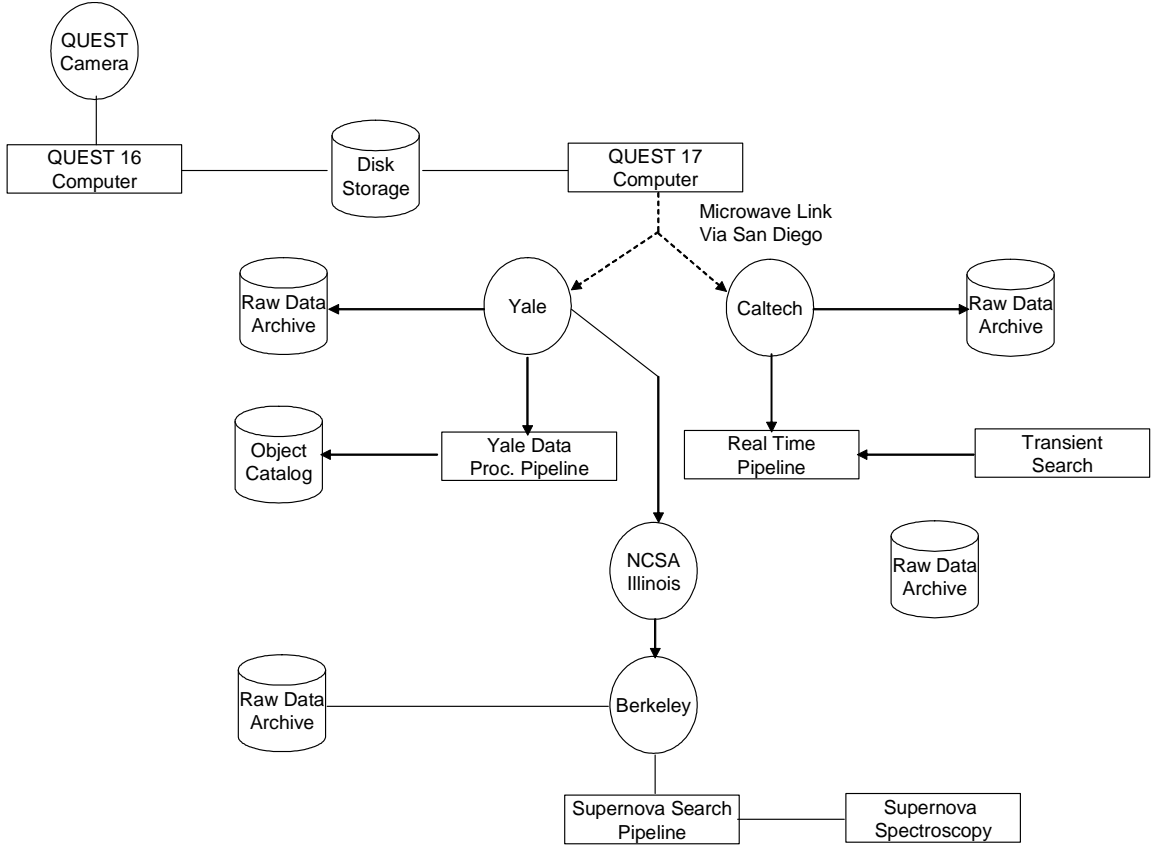

# The QUEST Data Processing Software Pipeline

Peter Andrews, Charles Baltay, Anne Bauer, Nancy Ellman,
Jonathan Jerke, Rochelle Lauer, and David Rabinowitz
(Yale University)

Julia Silge
(Quinnipiac University)

## ABSTRACT

A program that we call the QUEST Data Processing Software Pipeline has been written to process the large volumes of data produced by the QUEST camera on the Samuel Oschin Schmidt Telescope at the Palomar Observatory. The program carries out both aperture and PSF photometry, combines data from different repeated observations of the same portion of sky, and produces a Master Object Catalog. A rough calibration of the data is carried out. This program, as well as the calibration procedures and quality checks on the output are described.

Keywords:

Techniques: Photometric
Astronomical Data Bases: Miscellaneous
Survey
Astrometry

1. **Introduction**

The QUEST Large Area Camera, covering about 10 square degrees on the sky, has been installed and commissioned on the Samuel Oschin Schmidt Telescope at the Palomar Observatory. Science quality data taking with the camera started in the Fall of 2003. Since that time about 15,000 square degrees in an equatorial strip between declinations -25° to +25° have been observed multiple times with seven different color filters (Johnson U, B, R, I and Gunn r, i, and z).

This large data set is used to study a variety of science topics, and a number of different software packages are in use to reduce the data:

a)  The Yale Pipeline to analyze Drift Scan Data for quasar searches and strong lensing of quasars (Andrews 2003).

b)  The Caltech Data Cleaning Program to remove instrumental artifacts (Djorgovski et al., in preparation).

c)  The Berkeley Program developed by the SCP (Supernova Cosmology Project) to search for supernovae (Perlmutter et al. 1999).

d)  The Caltech Real Time Pipeline to carry out rapid, real time detection of transients (Djorgovski et al., in preparation).

The purpose of this paper is to present a more detailed description of the Yale Pipeline to process Drift Scan data.

A detailed description of the QUEST Large Area Camera has been presented in a previous paper (publication submitted to ….). A very brief summary of some of the features relevant to this paper is as follows: The camera consists of 112 CCDs arranged in 4 rows (labeled A, B, C, D) with 28 CCDs in each row (labeled 1 to 28), as shown in Figure 1. Each of the 4 rows has a different color filter; for example with Johnson filters rows A, B, C, D have R I B U filters, respectively. In the Drift Scan mode the telescope is locked in a fixed position typically for a whole night, and the rotation of the Earth causes the image of any given star to move across the camera. The camera is rotated in such a way that the images traverse rows A, B, C, D in sequence, thus providing data in four colors essentially simultaneously. The CCDs are 600 x 2400 pixels each, with a pixel size of 13 μ x 13μ (0.88 arc sec x 0.88 arc sec each). Near the equator it takes a star image 140 seconds to traverse each CCD, which is clocked synchronously with the motion of the star image. Given that there are 22 mm gaps between the rows it takes ~ 14 minutes to traverse the entire field of view of the camera. Some of the properties of the camera are summarized in Table 2.

The 28 CCDs in a row, which we call the 28 columns, are each at a different declination, and thus the star images move at different rates across the 28 CCDs. In drift scan mode the CCDs would have to be synchronized at slightly different clocking rates. There was a concern that this would introduce too much noise if one CCD were read out (typically microvolt signals) while a neighboring CCD was clocked (typically 10 volt signals). To avoid this, all CCDs are clocked synchronously at the same rate, but clocking signals are dropped at different intervals for the different columns to achieve the appropriate average clocking rates. These "line drops" have to be carefully taken into account in the data reduction programs.

## 2. General Description of the Yale Pipeline

The software packages to process drift scan data were written at Yale University. They were designed to deal optimally with closely separated sources (caused by gravitational lensing), to deal with the peculiarities of the data from the QUEST camera (such as dropped lines, etc.) and to process the huge volume of data in a time efficient manner. It was a priority to be able to process each night's data during the following day so as not to fall behind in the data reduction.

The first version of this pipeline was written (Andrews 2003) to analyze data from the QUEST1 survey using a 16 CCD prototype of the present camera on the Schmidt telescope at the CIDA Observatory in Venezuela. This first version was then modified and optimized for the Palomar QUEST survey.

The data from the camera is transferred from Palomar Mountain via a radio link and arrives at Yale in essentially real time. The data flow and archiving scheme is shown in Figure 2. The raw data from a night of drift scanning (a 4.6°wide in declination and typically 8 hours x 15°/hour = 120°long strip) is packaged into 28 folders, for the 28 columns of CCDs, each folder containing the full night's data stream in each of the 4 colors. The 28 columns are independent data sets each covering different declination strips of the sky, but the 4 CCDs in each column contain the same objects, in different colors. Initially each column was processed on a different processor, using a farm of 28 processors to simultaneously process the data. At the present time, with faster processors available, 12 processors are sufficient to process a full night's data every night. The raw data is compressed by a factor of two using a lossless compressions program and archived. A typical night's data is about 50 GBytes compressed.

The pipeline consists of three major programs:

a) The PHOTOMETRY program, carrying out both Aperture and PSF photometry.

b) The COMBINE program, which combines repeated measurements of any given object from observations on different nights and writes a Master Catalog of all objects detected.

c) The ANALYSIS program, which reads the Master Catalog, applies the photometric calibrations, and sets up the framework for users to carry out the science data analysis, generate distributions, and write various output files.

The components of the Pipeline are described in the following sections of this paper. The final section of the paper describes the photometric calibration of the data.

## 3. The PHOTOMETRY program

### 3.1 Data Formatting

Each column of data, spanning ~ 120° in RA, is saved on disk as images (frames) of 640 x 2400 pixels each. This is the size of the CCDs, and corresponds to 0.15° x 0.58° on the sky. A typical night's data thus consists of 28 columns about 200 frames long each.

### 3.2 Bias Subtraction and Flat Fielding

The next step is to carry out bias subtraction and flat fielding on a frame-by-frame basis. The flat fielding parameters are determined prior to running of the pipeline for each of the 112 CCDs in the camera using twilight flats taken periodically at the telescope.

### 3.3 Objection Detection

Object detection is carried out on one frame at a time, separately for each of the four colors. Since the sky level across a whole frame can vary, each frame is subdivided into thousands of small regions and the local sky level for each region is calculated as a two-stage clipped average of the pixels in that region. The object detection algorithm steps through the pixels in an image in a raster fashion and assembles objects out of pixels above a threshold of 2.5 standard deviations above sky background, An object is completed when it is surrounded on all sides by below threshold pixels. Objects are kept if they consist of at least four adjacent pixels above threshold. This routine also has a map of bad pixels or columns and these are taken into account in object finding.

### 3.4 CCD to CCD Coordinate Transformations

The same objects should be present in all four (different color filter) CCDs in each of the 28 columns. To relate the pixel locations of these to each other the pixel coordinate transformations between the relevant CCDs need to be calculated. The process starts with a triangle matching algorithm on the first frames. Since subsequent frames should to first order have the same transformations, the fits are extended to all frames of a drift scan to obtain a global set of transformations. Once this global linear transformation is found, nonlinear effects are taken into account. The most important of these is the effect of dropped parallel clock signals, as discussed above, used to correct for the differential star image drift speeds across different CCDs. These corrections, although very small, are nevertheless significant. The accuracy of the final CCD to CCD transformations has been measured to be 0.08 pixels in both the x and y directions on the CCDs. The final transformation is then used to match the objects detected in the four different colors to each other to make up the list of detected objects.

### 3.5 Astrometry

Up to this point transformations have been obtained to relate positions of object on one color CCD to the positions of these objects on CCDs with another color filter (chip to chip transformation). To obtain the transformation from the CCD position to actual astrometric positions on the sky the USNO A-2.0 catalog (Monet 1998) is used.

The header of each frame of data has the approximate location of the frame on the sky from the setting of the telescope during observations. Starting with this first approximation a triangle matching algorithm is used with selected bright objects to relate the pixel coordinate of each frame of data to the J2000 RA and Dec coordinates of stars on the USNO catalog. Small but significant effects, such as line drops discussed above, are taken into account. If the transformation for a frame fails to converge, the transformation is obtained by interpolating between adjacent frames.

It should be noted that the USNO catalog is in J2000 coordinates, which is not what is observed on any given night. The USNO coordinates are precessed to the epoch of the observations before the matching is attempted. Once the final transformations are obtained the astrometric coordinates of the drift scan survey are precessed back to J2000 coordinates.

To judge the precision of the astrometry of this procedure we compared the Right Ascension (RA) and Declination (Dec) from this survey with the astrometric standards of Stone et al. (Stone et al. 1999). We found that the mean values of USNO and the Stone Standards disagree by about 0.2 arc sec. To bring our data into agreement with the Stone, et al. standards we applied a final calibration correction of typically -0.1 arc sec in RA and -0.2 arc sec in Dec. After this final correction (which was applied in the ANALYSIS program described below) the comparison with the astrometric standards is shown in Figure 3. We see from this comparison that the astrometric precision of the drift scan data is 0.1 arc sec. However, over the entire area of our survey the systematic error is limited by that of the USNO catalog that we use for calibration.

## 3.6 The SEEING Routine

This routine takes the list of detected bright objects and for each determines the FWHM in each color of each frame. It then calculates the average seeing for each frame by performing a clipped mean of the measured FWHM values. These frame-wise computed values of FWHM, as well as the number of objects per frame, are stored for later use in the aperture photometry routine.

## 3.7 Aperture Photometry

Aperture photometry is performed in a standard way with four different apertures: one with one FWHM radius circle aperture, one with two FWHM radius, and two with fixed apertures with 3 and 10 pixel radii.

Typically the CCDs with different color filters have widely varying fluxes, with the R and I having the highest flux, B less, and U the least. Often objects are detected in the redder filter and not detected in the U filter. Even if an object is not detected in a particular color, a flux measurement is forced by using an aperture centered at the location obtained by transforming the centroid of the object detected in a brighter color. In fact we designate a lead color (Johnson R or Gunn r), obtain the coordinates of the centroid, and transform this coordinate to the other color CCDs. (The apertures used in the other colors are then centered on the transformed coordinates rather than the centroid determined with poorer statistics in the non-leading colors.) It was found that the transformations from CCD to CCD were sufficiently accurate that this procedure gave the best results.

For each aperture the flux is taken to be the total flux in the aperture minus the number of pixels times the average sky background per pixel. The average sky background is obtained as the clipped mean sky level in an annulus (typically of inside and outside radii of 20 and 25 pixels respectively) around the object.

At this point the flux is converted from Analog to Digital Converter Units (ADU's) to electrons using previously determined gains for each CCD. The error on the flux is calculated by adding the errors on the total flux and the error on the sky background in the aperture in quadrature. The flux and its error are then converted to the instrumental magnitudes using the formula

$$M_I = 25.0 - 2.5 \log (\text{flux in electrons})$$

For the one FWHM aperture, the ellipticity, orientation and size of the minor and major axes of the image, the FWHM, the sky level, the maximum pixel value and the number of pixels above threshold are recorded. Error flags are set to indicate problems such as saturated pixels, badly measured sky background, negative flux, etc.

## 3.8 PSF Photometry

In the aperture photometry routine described above, objects are detected as contiguous regions of pixels above some threshold, and the total flux is counted within an aperture centered upon each object.

This is fine for isolated stars, but it is easy to see that two stars of comparable magnitude and separated by 1 FWHM might be detected as one object using this method, and therefore the aperture would be placed somewhere between the two objects.

One of the main sources of contaminants in the early quasar candidate lists was multiple objects mis-identified as a single object, with the object center usually in between two stars. This easily led to a huge apparent variability for these objects, as the flux falling within with the misplaced aperture varied greatly with the seeing. In addition, the FWHM for the U filter is generally larger than other filters, which made these badly placed objects appear unnaturally U bright. Since variability and U band excess are our primary candidate selection mechanisms, it is easy to see that these types of objects were producing a significant number of false candidates. Even more worrisome, however, are the real quasar lenses which were falsely identified as a single object, consequently mis-measured, and lost from our candidate list.

The problem of detection and flux measurement for multiple objects can be solved properly if the Point Spread Function (PSF), or shape of the image is known. The PSF is used to separate a group of nearby objects by an iterative process of detection and fitting. Starting with a blob of flux, significant local maxima are located and each is called an object. Multiple PSFs are fit simultaneously to these groups of objects, and the result is subtracted out of the image. The region is then re-examined for significant positive residuals, which indicate an object which was previously hidden by brighter nearby stars. The process repeats so that any number of stars in a group can be pinpointed and measured.

The first step in this procedure is to obtain an accurate model of the PSF in the vicinity of the object of interest. The PSF is viewed as a two dimensional distribution of light impacting an imaging device which the atmosphere and the instrument produces from an idealized point source such as a distant star. We do not attempt to describe this as an analytic function but instead use the actual measurements of images of stars in the vicinity of the object of interest (a 20 pixel x 20 pixel area) to create a two dimensional table characterizing the PSF. This two dimensional distribution varies slightly depending on the location within the pixel that the image is centered on, i.e., the PSF is slightly different for a star centered at the center of a pixel then for a star centered near the corner of a pixel. Tot take this into account the central pixel is divided into 3 x 3 segments.

Following object detection, well measured point like images are selected (with FWHM consistent with a point source, good flux statistics but not saturated, etc) near the object of interest. These selected images are sorted into nine classes, each class consisting of images centered on one of the 3 x 3 segments of the central pixel. For each class the pixel fluxes of many images are summed and then normalized to a total unit flux. This produces nine different PSF tables. In the PSF fitting of a particular object the segment of the central pixel that the image is centered on is determined and the appropriate one of the nine PSF tables is used. This procedure has the advantage that local variations, irregularities, non circular symmetry, etc., of the PSF are properly taken into account.

The actual PSF fitting procedure is an iterative one. The fluxes in the individual pixels in the detected objects are examined. If there is a simple maximum the object is fit to a single PSF. If there are more than one local maxima the object is fit to a number of separate PSF's corresponding to the number of local maxima. After the first round of fits the fluxes expected from the fit PSF are subtracted from each image. If there is a significant positive residual, the residual flux is considered as a separate object and a new PSF is added centered on the residual flux. This allows the detection algorithm to detect multiple images or faint stars previously obscured by a close by bright star. The procedure is then repeated for a maximum number of iterations specified by the user.

The PSF fitting algorithm uses Powell's multidimensional methods, with Brent's inverse parabolic algorithm for each successive line minimization. During the detection stage the x,y and flux of all objects in a grouping are fix simultaneously to find the (hopefully) global minimum. For the later photometry stage only the fluxes are fit.

After the last round of PSF fitting the flux in each image and the error in that flux are obtained from the $\chi^2$ fitting procedure. This PSF fitting procedure turns out to be quite robust; it fails to give a meaningful flux less than 1% of the time. The fluxes are converted to instrumental magnitudes using the same formula as given above for aperture photometry.

The errors obtained from the $\chi^2$ fits are essentially statistical errors. To get a measure of systematic errors in this procedure we compared the results of repeated measurements of the same object in the same filter carried out at different nights. It was found that a systematic error of 0.015 magnitudes had to be added in quadrature to the statistical errors to bring the errors into agreement with the rms of the repeated measurements.

4. **The COMBINE Program**

In the PHOTOMETRY program described above the measurements of any given object on a particular night in the four different color filters are treated in a correlated fashion. The program processes one night's data at a time. In the Palomar-QUEST

Drift Scan Survey the same piece of sky has typically been observed 4 to 8 times during different nights. It is the purpose of the COMBINE Program to collect the results of the different observations of any given object. The output of the COMBINE Program is the Master Object Catalog containing for each object all of the information from all measurements in all colors and all of the different nights. This Master Catalog is updated periodically as the survey progresses.

The COMBINE Program consists of two parts, the LOAD section and the COMBINE section. The LOAD section takes the output of the PHOTOMETRY program and reformats the data into structures more convenient for further analysis. The PHOTOMETRY program output is organized by nights of observation. The LOAD output has the data organized into bins of 0.25 square degrees each on the sky (approximately 60,000 bins for the sky covered by the survey).

The COMBINE section works on one square degree bin of data at a time, stepping in turn through all of the bins. In each bin the routine looks at each entry in turn, using the RA and Dec coordinates to match up all of the multiple measurements of the same object. The matching algorithm used is the following, which works in two passes.

In the first pass, a separate list is generated for each scan covering the same area. Each list records the RA, Dec, and brightness measurements of all the objects detected by the corresponding scan. For each object in each list, the algorithm then checks the other objects in the same list and in the other lists to see if they are separated by $< 0.5"$. Such objects are grouped together as observations of the same master object. After the first pass, a list of master objects is created, each coupled to a group of one or more observations from one or more scans. During this first pass, however, occasional ambiguities arise. For example, if an extended object is resolved into two objects separated by $< 1"$, both components can match a single master object. Also, two close objects may be separated in one scan but not in another. To handle these cases, any multiple matches from a given scan to a single master object are provisionally listed as separate master objects. In the second pass, all close master objects are re-examined. If an observation associated with one master object is better grouped with those of another, the new association is made.

For objects near the edge of a 0.25 square degree zone, data from the neighboring bin is also examined for possible matches.

The output of the COMBINE Program, the Master Catalog, is a catalog of all objects detected one or more times, with objects organized into the 0.25 square degree bins, and within each bin listed sequentially by RA. The information on all of the multiple measurements of each object is not physically copied into the Master Catalog; instead, the Master Catalog, in addition to some summary information such as RA, Dec, number of measurements in each color, etc., contains pointers to where the individual measurements can be found in the LOAD output.

## 5. The ANALYSIS Program

The ANALYSIS Program is the last section of the Yale Pipeline. This program reads the Master Catalog one object at a time and performs the following functions:

a) The pointers are followed to the LOAD output and all of the relevant information for all of the measurements of this object are collected.

b) The photometric calibrations are applied to each individual measurement to convert the instrumental magnitudes to calibrated magnitudes. The appropriate calibration parameters have been obtained by a previous calibration procedure and are in the ANALYSIS Program. The details of the calibration procedure are discussed in Section 6 of this paper.

c) Data quality cuts are applied to flag individual measurements that are of inferior quality and should not be used in calculating average magnitudes and other science analysis. These quality cuts set limits on quantities like sky brightness, extinction, seeing quality, etc. (see section 7 for the actual cuts used).

d) A framework is provided to carry out further science analysis of the data and to facilitate writing output text, graphs, catalogs, etc., as desired.

## 6. Photometric Calibration

The fluxes for detected objects recorded by the camera have to be corrected, or calibrated, to account for two different kinds of effects. The first is extinction due to light cloud cover or other atmospheric effects that tend to vary not only from night to night but possibly more frequently during one night of observation. The second is due to the non uniform response between CCDs, and even variation of response over the area of one CCD, filter transmissions, non linear response of the CCDs, and other instrumental effects. These second kind of instrumental effects do not vary significantly with time (they are constant to better than the 5% level) except for discrete incidents like replacement or adjustments of electronics, telescope mirror resurfacing, etc., which are carefully noted and kept track of. Thus the photometric calibrations consist of two parts: the extinction corrections, and the calibration of the instrument.

### 6.1. The Extinction Corrections

If all nights were perfectly photometric, and all observations were made at the same air mass, we would not need this correction. On nights with a significant cloud cover or other adverse weather conditions, the dome does not open and no observations are made. However, even on nights when data are recorded, there is often a significant amount of extinction. In this survey we also use the non-photometric nights with extinctions less than 0.5 magnitudes. We thus have to make a correction for the extinction. The only large area survey that covers all of the area of the Palomar-QUEST survey is the USNO survey (Monet 1998). We started out using the USNO catalog for the extinction correction but found that the photometry of USNO was sufficiently variable not to be useful for this purpose. We therefore adopted the following procedure using only our own survey data, making use of the fact that we covered each area of sky in the survey multiple times during different nights.

The procedure starts by taking each drift scan strip (recall that these are 4.6° wide in declination) and dividing it into 1° segments in right ascension (about 4 minutes of scan time). In each color filter all the observations in a given RA segment were collected and the brightness of a sample of stars was compared. We then assume that the observation with the highest flux for the selected stars was a good approximation to photometric and assign an extinction correction to all of the other observations to bring them up to the level of the brightest night. The distribution in the extinction corrections (in magnitudes) for the 1° RA segments is shown in Figure 4. We see from this Figure that ~ 80% of the data is within 20% of photometric. If the extinction correction is larger than 0.5 mags, the data are flagged as bad by the ANALYSIS Program and not used for science data analysis. About 10% of the data are lost due to this cut. These extinction corrections were fed into the ANALYSIS Program described above and were used to correct the instrumental magnitudes. The precision of these correction are discussed in section 6.3 below.

**6.2 Calibration of Instrumental Effects**

The Sloan Digital Sky Survey (SDSS) Data Release 4 catalog (Adelman-McCarthy et al. 2006) is used to do the photometric calibration of the QUEST survey, using areas of sky where the two surveys overlap. Several clear nights of QUEST data, both with Johnson and with Gunn filters, are selected. Objects in QUEST are matched to those in SDSS (using objects classified as stars by SDSS) and the QUEST instrumental magnitudes (after being corrected for extinction as described above) are compared to the SDSS quoted magnitudes. The calibration consists of calculating the Zero Point Offset, i.e., the magnitude that has to be added to the QUEST instrumental magnitudes to bring them into agreement with SDSS. This is straightforward for the Gunn filters since those are the filters used by SDSS. To calibrate the data with Johnson filters an algorithm (Fukugita et al. 1996) was used to convert the Gunn colors of the SDSS catalog to Johnson colors with sufficient precision to compare to QUEST data with Johnson filters.

This calibration procedure was carried out for all 112 CCDs, separately with the Johnson and Gunn filters, so there were a total of 224 calibrations. Furthermore, the calibration of a single CCD was allowed to vary as a function of the column across the CCD, the magnitude of the object, and the sky brightness. The full calibration of the camera thus consisted of 112 CCDs x 2 colors x 10 regions across a CCD x 6 magnitude bins x 6 sky brightness bins = 80, 640 calibration parameters. Thanks to the substantial overlap area with the SDSS survey we could use ~ $10^6$ stars to evaluate all of these parameters. These parameters were fed into the ANALYSIS Program described above and were used to transform the instrumental magnitudes (already corrected for extinction) to obtain the final calibrated magnitudes.

### 6.3 Quality of the Photometric Calibration

Extensive checks were carried out to check the quality of the photometric calibration. One of the most stringent tests is to look at the absolute photometric magnitude calibration by comparing with the SDSS data using a 1500 square degree region where the two surveys overlap. We compared each of the eight color filters separately. The comparison for the Gunn r filter is shown in Figure 5. We use in this comparison the entire QUEST data set data from September 2003 to September 2006, with the quality cuts applied in the ANALYSIS Program, including both photometric and non photometric nights. The comparison, Figure 5a, has a standard deviation $\sigma = 0.11$ mags. This includes both the SDSS and QUEST statistical errors as well as the calibration errors. To separate the two we plot the distribution of the magnitude difference between SDSS and QUEST divided by the total error, where the total errors is the SDSS and QUEST errors on each object added in quadrature to the estimated calibration error. We vary the calibration error estimate and choose the one that gives a $\sigma = 1$ for this distribution. For the r filter, the best estimate of the overall systematic and calibration error was 0.08 magnitudes, as shown in Figure 5b. The results for the estimated calibration error for all of the other color filters including both photometric and non photometric nights vary between 0.06 and 0.11 mags except for the U and B filters where it is more like twice this. There is also typically 15% of the data in a non Gaussian tail.

This rough overall photometric calibration error is for the entire survey, including non-photometric nights. If we restricted the comparison to photometric nights only, the calibration errors would be considerably smaller. Furthermore this procedure uses a single set of magnitude zero points for all nights, with the relative extinction corrections as described in section 6.1. No attempt was made at this point to determine color terms in the corrections. In any particular science analysis where better photometric precision is required calibration procedures appropriate for that particular analysis will have to be used.

For example, to check the systematic errors of the relative magnitude calibration we took frames with repeated measurements of the same area of the sky spread out over the three years of the survey. We normalized the different epochs to each other using a sample 10 or more field stars, and plotted the distribution of the deviation of the magnitudes of the individual measurements from the average magnitude. An example for such a plot for the Johnson R filter, using data for bright stars with six or more measurements at different epochs, is shown in Figure 6a. The standard deviation of this distribution is $\sigma = 0.03$ mags. This error includes both the statistical and the systematic components of the error. To separate the two we used a procedure similar to that described above. We divided each deviation from the average by the error, where the error was taken as the statistical and estimated systematic errors added in quadrature, and varied the estimated error. We needed a systematic error of 0.02 mags to produce a distribution with $\sigma = 1$ as shown in Figure 6b. Thus our best estimate for the systematic error on relative photometry is 2%. It is also gratifying that a single Gaussian with $\sigma = 1$ is a good fit to the data with very little non Gaussian tail.

7. **Completeness and Purity of the Data Sample**

   To check the overall quality of the Yale data processing pipeline, we carried out further comparisons with the SDSS Data Release 4 (Adelman-McCarthy et al. 2006) in the area where the two surveys overlap.

   To estimate the completeness of the Palomar-QUEST (PQ) data set we selected all objects in the SDSS Catalog in the PQ SDSS overlap areas and looked for a match in the PQ data within two arc seconds in RA and Dec. The completeness of PQ was taken to be the fraction of the SDSS objects that were found in PQ. This completeness, as a function of SDSS r magnitudes, is shown as the upper solid line in Figure 7.

   We see from this plot that the overall completeness of the Palomar QUEST data set is around 97% up to r magnitude of 21, and falls to 50% around mag 22.

   To check the purity of the PQ data set we took objects in the PQ Master Catalog in the PQ – SDSS overlap area and asked what fraction of the PQ objects have a match (within two arc seconds) in the SDSS Catalog. For this comparison the standard quality (QA) cuts were made on PQ objects in the ANALYSIS program:

   - Astrometry residuals $\leq$ 1 arc sec
   - Sky level $\leq$ 6 times normal
   - Extinction correction $\leq$ 0.5 mags
   - Image FWHM $\leq$ 4 arc sec
   - Objects were removed if the image was saturated, near the edge of the CCD, or had a PSF fitting error flag

   Two further selection criteria were used to eliminate false detections:

   a) To remove cosmic rays, only objects were kept with either detections in two different filters of one night, or detections in a single filter but in at least two different nights.

   b) To remove fake detections near saturated bright stars, detections near bright USNO stars and detections with more than 10 neighboring detections in a 10 arc second radius, were removed.

After these selection criteria were applied, the purity of the PQ sample was found to be around 98% up to r magnitude 21, as shown in Figure 8.

The quality cuts and selection criteria used to clean up the PQ data set, as described above, reduce the completeness of the data set (some legitimate objects are close to bright stars, or are lost due to high sky background due to the moon or large extinction due to clouds, etc.). The completeness with all of the selection criteria applied is shown as the lower, dashed curve on Figure 7. It should be pointed out, however, that for any particular science analysis the severity of these selection criteria can be varied, thus optimizing purity at the expense of completeness or vice versa.

8. Conclusion

The Yale data processing pipeline described in this paper has been in place essentially in its present final form since the beginning of science quality data taking of the Palomar QUEST survey in the fall of 2003. The pipeline has been run regularly every day during drift scan data taking and has kept up with the data on a daily basis. The program has not been changed in any significant way since the beginning to ensure a uniform quality to the survey data sample. The calibration procedures described in this paper have been carried out after data taking has begun and these calibrations have been completed just recently.

9. Acknowledgements

This work has been supported by the Department of Energy Grant No. DEFG-02-92ER40704 and the National Science Foundation Grant No. AST-0407297.

Figure: 1  Arrangement of the 112 CCD's in the QUEST Large Area Camera

Figure 2 Palomar QUEST Data Distribution, Processing and Archiving

Figure 3:  Astrometric precision in the Drift Scan Mode.  The figures show the agreement of the QUEST Right Ascension and Declination measurements with the astrometric standards of Stone, et al Stone et al. 1999)

Figure 4:  Distribution of the Extinction Corrections for 1 Degree Sections of Right Ascension

Figure 5:  Comparison of the calibrated QUEST magnitudes in the Gunn r filter to the SDSS data for the total QUEST/SDSS overlap area (~1550 square degrees) and including all QUEST data from the first 3 years of the survey.

a) Distribution in the SDSS-QUEST magnitude difference.  The standard deviation is $\sigma=$ 0.11 mags which includes both the statistical errors and the calibration errors.
b) Distribution in the SDSS-QUEST magnitudes divided by the combined errors after adding an 8% calibration error in quadrature to the SDSS and QUEST statistical error, which is required to give a Gaussian with a standard deviation of 1.

Figure 6:  a)  Distribution in the deviation of the magnitudes for individual measurements from the average magnitude for multiple measurements of the same object at different epochs.
c) Distribution in the deviation of individual measurements from the average, divided by the total error, where the total error is the statistical error on each measurement added in quadrature to a 0.02 mag systematic error, to produce a Gaussian distribution with a $\sigma=1$.

Figure 7:  Completeness of the Palomar-QUEST data set as compared to the SDSS survey. The upper solid line is without any cuts, the lower dashed line is with the selection criteria described in the text.

Figure 8:  Purity of the Palomar-QUEST data set with the quality cuts described in the text, as compared to the SDSS survey.

TABLE 1
Properties of the QUEST Large Area Camera
______________________________________________

| Property | Value |
|---|---|
| Number of CCDs ……………………….. | 112 |
| Array Size, CCDs ……………………. | 4 x 28 |
| Array Size, Pixels …………………….. | 9600 x 16,800 |
| Array Size, cm ………………………… | 19.4 cm x 25.0 cm |
| Array Size on Sky …………………….. | 3.6° x 4.6° |
| Sensitive Area ………………………… | 9.6 square degrees |
| Total Pixels …………………………… | 161 x $10^6$ |
| For each CCD: | |
|     Pixel Size ……………………. | 13 μ x 13 μ |
|     No of Pixels …………………. | 600 x 2400 |
|     Pixel Size on Sky ……………. | 0.88" x 0.88" |
| Read noise ……………………………. | 6 electrons |
| Dark Current | |
|     At room temperature………… | 500 pA/pixel |
|     At -100°C ……………..…… | 0.1 electrons/sec |
| Quantum Efficiency …………….…… | 95% at Peak of 6000 A |
| Antireflection Coating ………….…… | 450 nm thick Bismuth Oxide |

______________________________________________

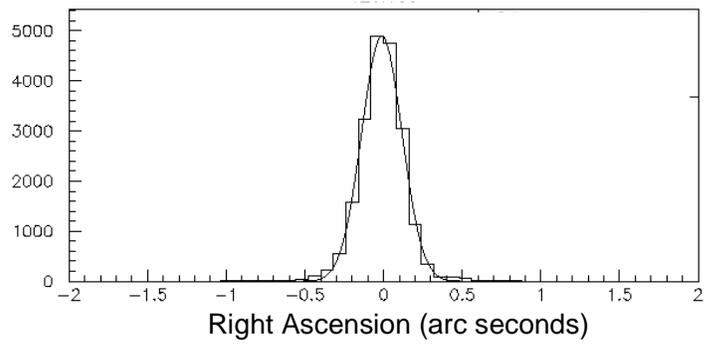
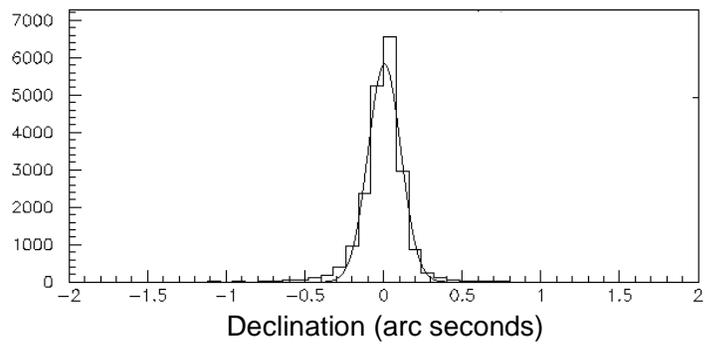

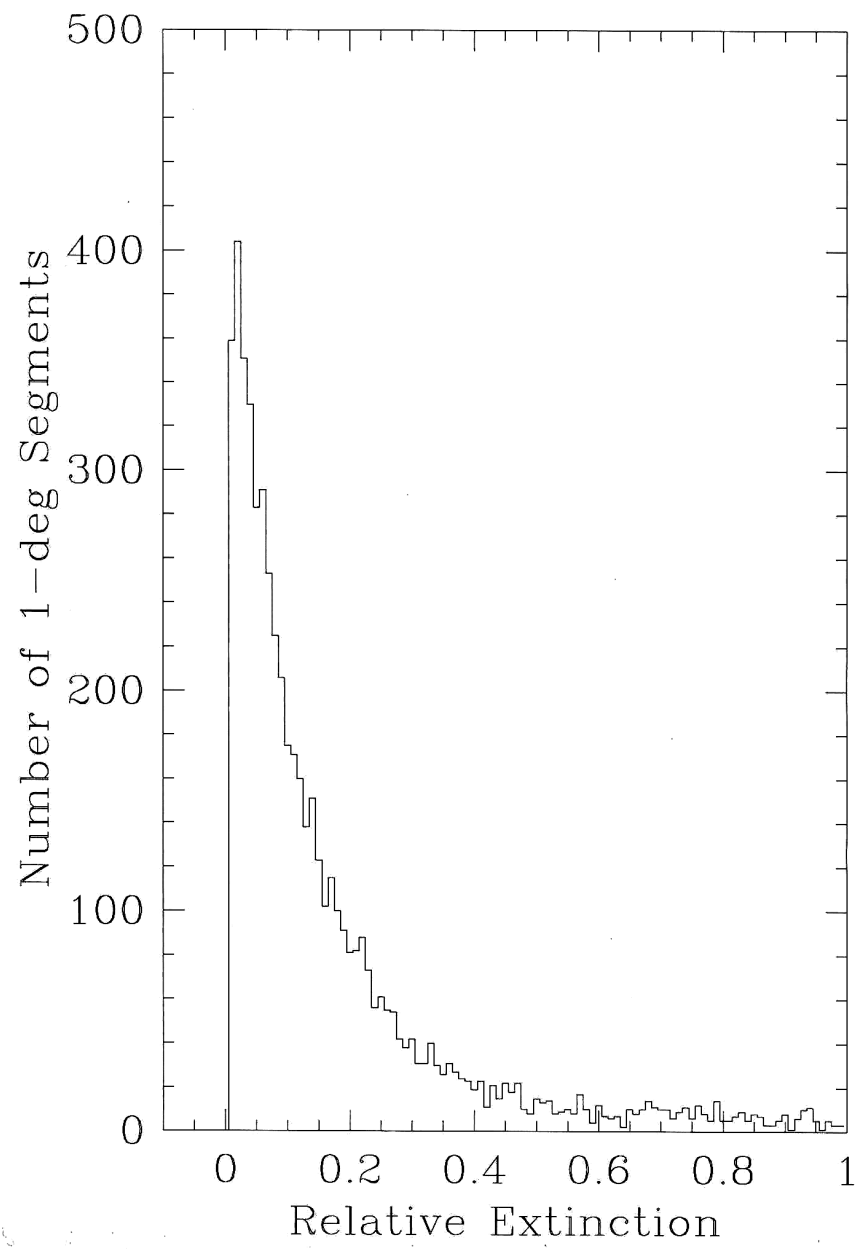

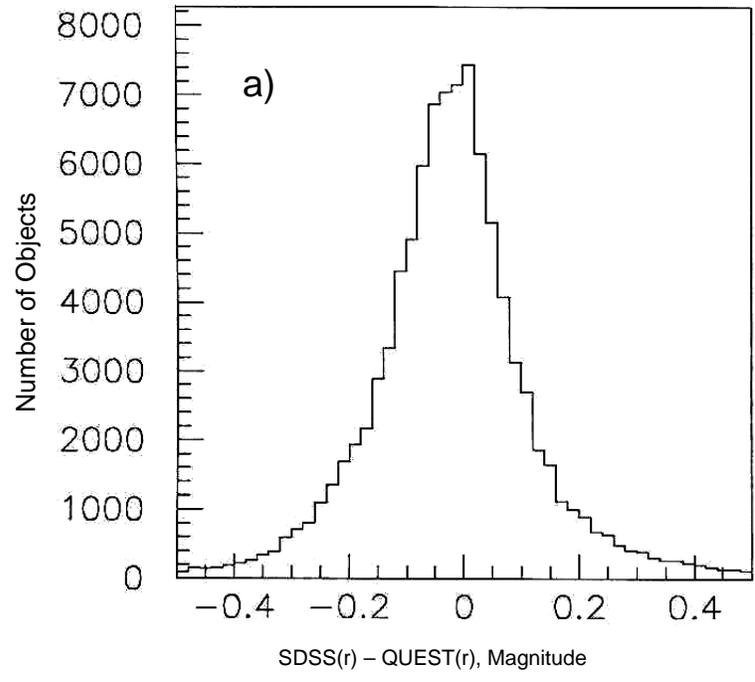

SDSS(r) − QUEST(r), Magnitude

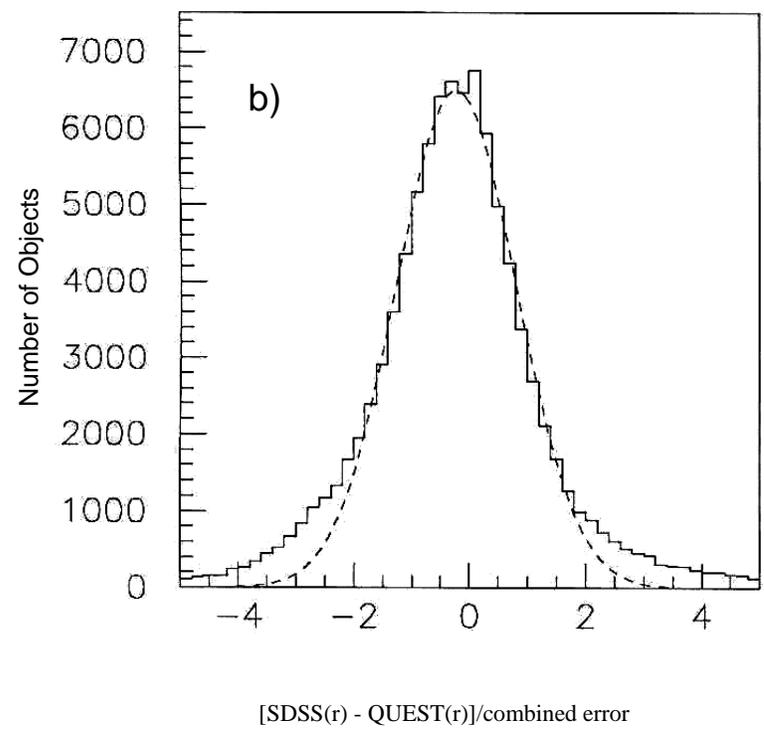

[SDSS(r) - QUEST(r)]/combined error

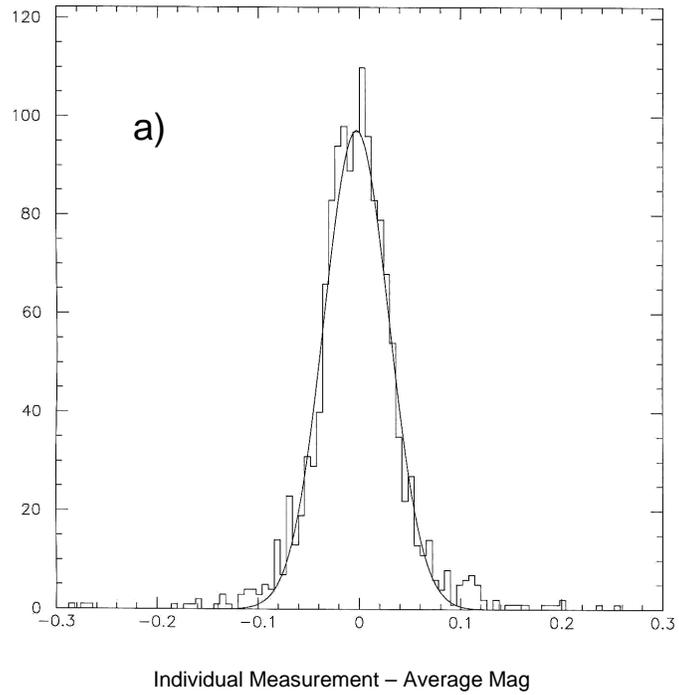

Individual Measurement – Average Mag

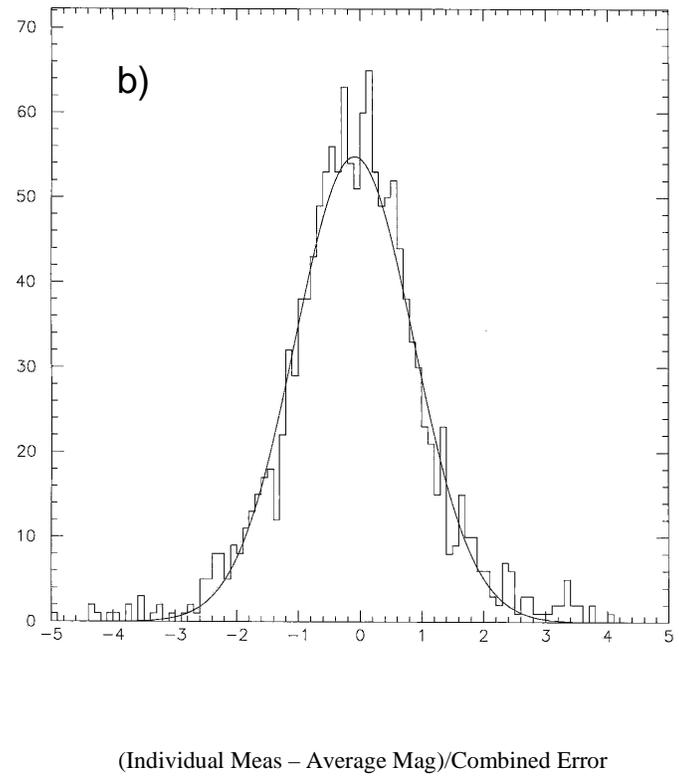

(Individual Meas – Average Mag)/Combined Error

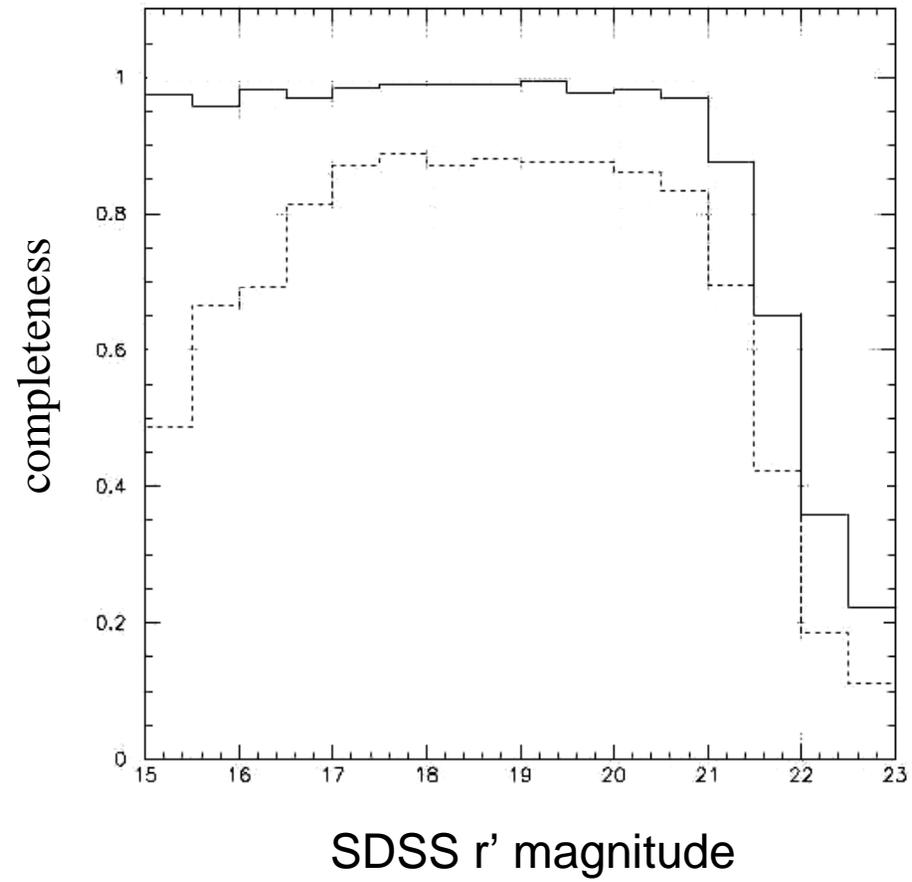

SDSS r' magnitude

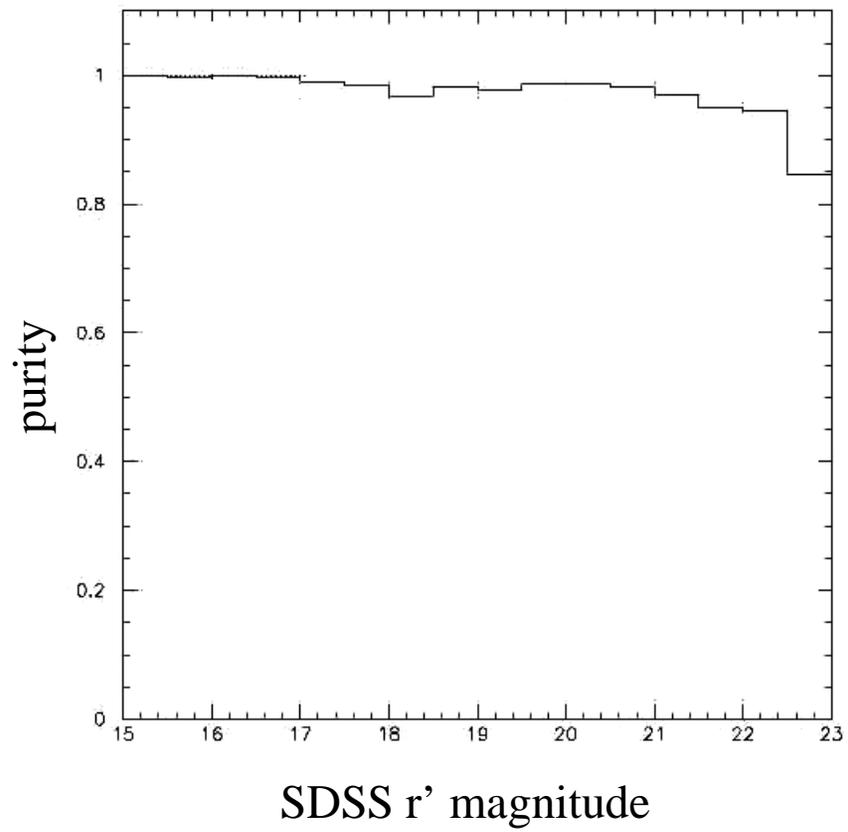